\documentclass[twocolumn,secnumarabic,amssymb,nobibnotes,aps,prb]{revtex4-1}
\usepackage{color}
\usepackage{graphicx}
\usepackage{natbib}
\usepackage{url}
\usepackage{textcase}
\usepackage{bm}
\usepackage{amsmath} 
\usepackage{ulem} 
\begin{document}

\title{Mott metal-insulator transition induced by utilizing a glass-like structural ordering in low-dimensional molecular conductors}

\author{Benedikt Hartmann}
\email[Email: ]{hartmann@physik.uni-frankfurt.de}
\affiliation{Institute of Physics, Goethe-University Frankfurt, 60438 Frankfurt (M), SFB/TR49, Germany}
\author{Takahiko Sasaki}
\affiliation{Institute for Materials Research, Tohoku University, Sendai, 980-8577, Japan}
\author{Jens M\"uller}
\affiliation{Institute of Physics, Goethe-University Frankfurt, 60438 Frankfurt (M), SFB/TR49, Germany}

\date{\today}

\begin{abstract}
We utilize a glass-like structural transition in order to induce a Mott metal-insulator transition in the quasi-two-dimensional organic charge-transfer salt $\kappa$-(BEDT-TTF)$_2$Cu[N(CN)$_2$]Br. In this material, the terminal ethylene groups of the BEDT-TTF molecules can adopt two different structural orientations within the crystal structure, namely eclipsed (E) and staggered (S) with the relative orientation of the outer C--C bonds being parallel and canted, respectively. These two conformations are thermally disordered at room temperature and undergo a glass-like ordering transition at $T_g \sim 75$\,K. When cooling through $T_g$, a small fraction that depends on the cooling rate remains frozen in the S configuration, which is of slightly higher energy, corresponding to a controllable degree of structural disorder. 
We demonstrate that, when thermally coupled to a low-temperature heat bath, a pulsed heating current through the sample causes a very fast relaxation with cooling rates at $T_g$ of the order of several 1000\,K/min. The freezing of the structural degrees of freedom causes a decrease of the electronic bandwidth $W$ with increasing cooling rate, and hence a Mott metal-insulator transition as the system crosses the critical ratio $(W/U)_{c}$ of bandwidth to on-site Coulomb repulsion $U$. Due to the glassy character of the transition, the effect is persistent below $T_g$ and can be reversibly repeated by melting the frozen configuration upon warming above $T_g$. Both by exploiting the characteristics of slowly-changing relaxation times close to this temperature and by controlling the heating power, the materials can be fine-tuned across the Mott transition. A simple model allows for an estimate of the energy difference between the E and S state as well as the accompanying degree of frozen disorder in the population of the two orientations. 
\end{abstract}

\maketitle

\section{Introduction}
\begin{figure*}[htb]
\includegraphics[width=\textwidth]{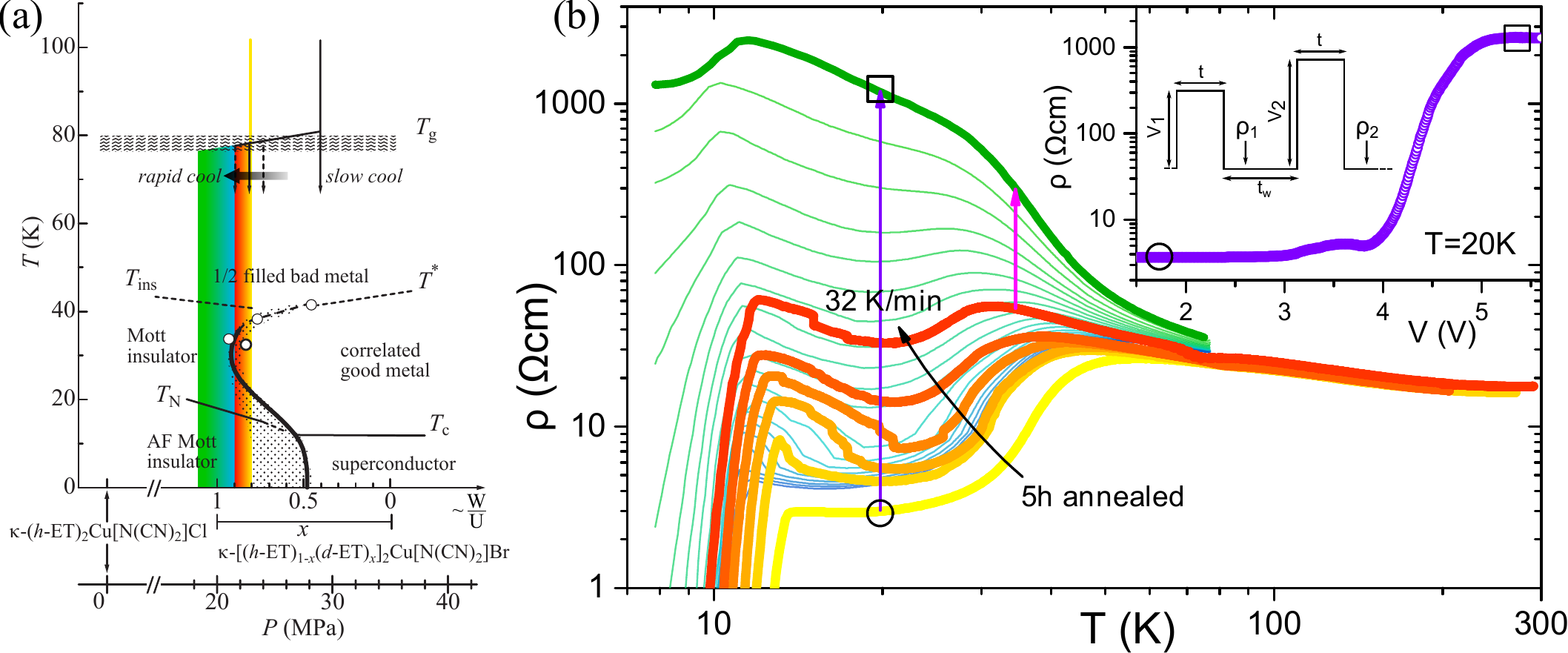} 
\caption{(Color online.) (a) Phase diagram of $\kappa$-(ET)$_2X$. The horizontal axis represents the substitution ratio $x$ for $\kappa$-[($h$-ET)$_{1-x}$($d$-ET)$_{x}$]$_2$Cu[N(CN)$_2$]Br, and the corresponding hydrostatic pressure for $\kappa$-(ET)$_2$Cu[N(CN)$_2$]Cl. Open circles indicate values taken from the literature of the second-order critical endpoint of the first-order Mott MIT (thick solid curve) in $\kappa$-(ET)$_2$Cu[N(CN)$_2$]Cl, see \cite{Sasaki2005} and references therein. 
The horizontal position also changes with the cooling rate employed at $T_g \simeq 75 - 80$\,K. Colors indicate positions after different preparation conditions in this work. Yellow marks the pristine (slowly cooled) state.
(b) Temperature dependence of the resistivity, $\rho$ {\it vs.} $T$, of $\kappa$-[($h$-ET)$_{0.2}$($d$-ET)$_{0.8}$]$_2$Cu[N(CN)$_2$]Br for cooling rates ranging from 5h annealing at 75\,K (yellow) to $q = 1$, 2, 5, 10, and 32\,K/min (red). Vertical arrows indicate the possibility to reach an insulating state (thick green line) after applying a voltage pulse $V$ at different temperatures. Intermediate states have been realized by relaxing the system at $75$\,K for $\sim 10$\,min and subsequent cooling (green to blue). Inset shows a discrete $\rho_n$ {\it vs.} $V$ measurement (sketched voltage-pulse sequence with $t = 2$\,s and $t_{\rm w} = 20$\,s is explained in the text) at $T = 20$\,K (compare violet arrow).}
\label{fig:01}
\end{figure*}
The family of organic charge transfer salts $\kappa$-(BEDT-TTF)$_2$X, where BEDT-TTF (representing C$_6$S$_8$[(CH$_2$)$_2$]$_2$, commonly abbreviated as ET) denotes bis-ethylenedithio-tetrathiafulvalene and X is a polymeric anion, are among the rare examples for a purely bandwidth-driven Mott metal-insulator transition (MIT) \cite{Kanoda1997}, a key phenomenon in modern condensed matter physics, where a gap in the charge-carrying excitations opens due to the Coulomb interaction between the electrons \cite{Imada1998}. The observed rich phenomenology of ground states in these materials \cite{Toyota2007,Lebed2008,Powell2011} is due to both the highly tunable nature of the correlation strength of the charge carriers and the strong coupling of the latter to intra- and intermolecular vibrational modes of the underlying crystal lattice. The low energy scales thereby realized in these materials and the possibility to fine-tune their position in the generalized phase diagram makes them unprecedented model systems for studying the Mott MIT in reduced dimensions.\\ 
The Mott MIT can be accessed, {\it e.g.}, by applying a moderate pressure of 300\,bar to the antiferromagnetic insulator $\kappa$-(ET)$_2$Cu[N(CN)$_2$]Cl (denoted $\kappa$-Cl), which shifts the system to a superconducting metallic state, see the phase diagram in Fig.\,\ref{fig:01}(a). A corresponding change in the ratio of bandwidth to on-site Coulomb repulsion $W/U$ \cite{Kanoda1997} can be achieved similarly by changing the anion to X = Cu[N(CN)$_2$]Br ($\kappa$-Br in short), which is a metal and superconductor at ambient pressure. Replacing the hydrogen atoms in the ET molecules' terminal ethylene groups [(CH$_2$)$_2$]$_2$ (ethylene endgroups, in short EEG) by deuterium in turn results in a smaller bandwidth $W$, which shifts the system's position from metallic towards the insulating side of the Mott MIT. Further fine-tuning is possible by varying the substitution ratio $x$ between the hydrogenated ET molecule ($h$-ET) and the deuterated one ($d$-ET) in $\kappa$-[($h$-ET)$_{1-x}$($d$-ET)$_{x}$]$_2$Cu[N(CN)$_2$]Br \cite{Yoneyama2004,Sasaki2005}, see Fig.\,\ref{fig:01}.\\
Besides the possibility to tune the interaction strength for studying fundamental aspects of the Mott transition, an aspect of considerable recent theoretical and experimental interest in this field of research relates to the interplay of correlation effects and disorder (Mott-Anderson scenario), see {\it e.g.}\ \cite{Analytis2006,Powell2009,Shinaoka2009a,Sano2010,Sasaki2012}.

Of particular importance for both tuning the correlation strength and considering disorder, are the vibrational degrees of freedom of the ET molecules' EEG, which have two possible conformations, where their relative orientation can be either parallel (eclipsed, E) or canted (staggered, S). Their vibrational properties and the relative population of the E and S states have been shown to have a strong effect both on the scattering of charge carriers and the electronic ground state properties \cite{Su1998,Su1998a,Tanatar1999,Stalcup1999}. For a system located close to the critical region $(W/U)_{c}$, the E and S population can even determine whether the ground state is insulating/magnetic or metallic/superconducting \cite{Kawamoto1997,Taniguchi2003,Sasaki2005,Taylor2008}.  
For $\kappa$-Cl and $\kappa$-Br, 
the population of the E and S states is thermally disordered at room temperature (see below), and the EEG system gradually orders upon cooling, favoring the eclipsed configuration due to its slightly lower energy \cite{Ishiguro1998}. The EEG ordering process, however, cannot be completed for kinetic reasons, since upon lowering the temperature their molecular motion/rotation slows down so rapidly that thermodynamic equilibrium cannot be reached and a short-range structural order becomes frozen-in. This is a glass-like transition \cite{Saito1999,Akutsu2000,Mueller2002,Mueller2004}, and the degree of EEG order/disorder depends on the glass transition temperature $T_g$, which in turn varies with the cooling rate $q = {\rm d}T/{\rm d}t$ such that faster cooling results in a higher $T_g$ and a larger degree of frozen-in disorder on the population of the E and S states \cite{comment}, see \cite{Toyota2007} for an overview.\\ 
The effect of the EEG glass-like ordering transition on the electronic (ground-state) properties is twofold: (i) The strong vibrations lead to an enhanced scattering contribution at elevated temperatures above $T_g$ \cite{Mueller2009,Brandenburg2012} and their frozen-in configuration below $T_g$ causes a random lattice potential, which gives rise to an additional contribution to the residual resistivity and affects the superconducting state and transition temperature \cite{Su1998,Su1998a,Yoneyama2004,Powell2004}. 
(ii) 
The anisotropic change of in-plane lattice parameters at $T_g$ results in a smaller ratio $t_{\rm inter}/t_{\rm intra}$ of the inter- and intra-dimer transfer integrals for more rapid cooling, which in turn leads to smaller $W/U$ \cite{Sasaki2005}, {\it i.e.}, the strength of electronic correlations can be controlled by varying the cooling rate through $T_g$. 

Although this cooling-rate effect on the bandwidth has been recognized as a highly useful tool for tuning the ground state properties and therefore has been frequently used, quantitative information on the E/S population and a systematic understanding of the interplay of bandwidth variation and frozen randomness are still lacking.\\ 
In this work, the glass-like structural transition is realized by applying a pulsed heating current to samples that are thermally coupled to a low-temperature heat bath. By this means a system located on the metallic side in the vicinity of the Mott critical region $(W/U)_{c}$ can be tuned from a metallic/superconducting to an insulating behavior with a resistivity change in the normal state of almost three orders of magnitude. A detailed analysis of thereby generated temperature quenches reveals cooling rates of the order of several 1000\,K/min. A simple model shows that even for such extreme conditions the relative amount of disordered ET molecules is smaller than 6\%. The ratio $W/U$ can therefore be precisely tuned in the presence of a more-or-less weak random lattice potential.
 
\section{Experiments} 
Single crystals of $\kappa$-Br, as well as the partially deuterated variant with $x = 0.8$, $\kappa$-[($h$-ET)$_{0.2}$($d$-ET)$_{0.8}$]$_2$Cu[N(CN)$_2$]Br (denoted as $\kappa$-$d_{0.8}$-Br), of cuboid-shaped morphology were grown by electrochemical crystallization as described elsewhere \cite{Wang1990,Yoneyama2004}. Resistance measurements were carried out in standard four-terminal geometry with current applied perpendicular to the highly-conducting planes using a low-frequency ac lock-in technique (Stanford Research 830). Voltage heat-pulses of variable amplitude and length have been applied using a dc source-meter (Keithley 2400). Further details of the measuring processes, as far as necessary, are given in the respective sections. Temperatures from $2 - 300$\,K were controlled in a VTI cryostat (Oxford Instruments) allowing for a maximum sample cooling rate of $q \gtrsim 30$\,K/min (sample not immersed in liquid Helium).

\section{Results and Discussion}
\subsection{Cooling-rate-dependent resistivity}
Figure\,\ref{fig:01}(b) shows the resistivity of $\kappa$-$d_{0.8}$-Br as a function of temperature for different cooling rates with 5\,h annealing at $T = 75$\,K (yellow curve, showing essentially metallic behavior) and continuous cooling through $T_g$ ranging from $q = 1$\,K/min, 2\,K/min, 5\,K/min and 10\,K/min to $q = 32$\,K/min (red curve). The latter curves show the characteristic re-entrant behavior, where the S-shaped MIT line is crossed twice. The above-described increase in the ratio of S/E population of the EEG with increasing cooling rate corresponds to a small expansion of the low-temperature in-plane lattice constants and therefore 
the systematic change in resistivity (black arrow) can be interpreted as a decrease of $W/U$ \cite{Yoneyama2004,Sasaki2005}, which is represented by a slight shift in the phase diagram, see colored lines in Fig.\,\ref{fig:01}(a). Here, we note that in recent quantum oscillation measurements of the related compound $\kappa$-(ET)$_2$Cu(NCS)$_2$ it has been shown that the disorder introduced by the partial substitution of ET molecules with their deuterated analogues does not cause additional quasiparticle scattering contributions \cite{Sasaki2011}. Likewise, for $\kappa$-[($h$-ET)$_{1-x}$($d$-ET)$_{x}$]$_2$Cu[N(CN)$_2$]Br investigated here, the effect of disorder by substitution is negligible, 
as demonstrated 
by an initial smooth increase of the superconducting $T_{\rm c}$ due to the chemically induced pressure effect \cite{Yoneyama2004}. 
%
\subsection{MIT induced by a single voltage pulse} 
Furthermore, an insulating state with an almost three orders of magnitudes higher resistivity (thick green curve) compared to the metallic state (yellow curve) was reached after a voltage pulse of 5\,V amplitude and $2$\,s duration was applied. The insulating state is persistent (confirmed for times exceeding 18\,h) and can be reached independent of the initial state, {\it i.e.}\ by applying the pulse at different temperatures, as indicated by the vertical arrows in Fig.\,\ref{fig:01}(b). Thus, the system can be permanently transformed from the superconducting into an insulating ground state. 
The system can be re-set into the metallic state by warming the sample above $T_g \sim 75$\,K and subsequent slow cooling thereby relaxing its resistivity. 
This phenomenology suggests that during the voltage pulse the Joule heating is sufficient to increase the sample temperature above the glass-like transition temperature $T_g$ and that after switching off the voltage, the coupling to the low-temperature heat bath results in a very rapid cooling through $T_g$. Subsequent measurements reproduce the results, and we find no signs of sample deterioration, such as micro cracks, which would permanently alter the resistivity profile. 
Starting from the highly resistive state, we can then utilize the glass-like relaxation which is accessible in the temperature range around $T_g$ of $70 - 80$\,K by annealing in steps of $\sim 10$\,min in order to intentionally create intermediate states with reduced resistivity in subsequent cool-down measurements, see
Fig.\,\ref{fig:01}(b), thinner curves from green to blue. In this context, it is interesting to recall the scaling analysis of the conductance of $\kappa$-Cl tuned by pressure aiming to determine the critical exponents of the Mott transition \cite{Kagawa2005}. A similar analysis of the present data seems difficult at first view, since the parameters cooling rate $q(T_g)$ or, likewise, relaxation time at $T_g$ do not necessarily scale linearly with pressure, which is the relevant thermodynamic variable. As we will show below, however, a simple model allows for assigning the occupation ratio of eclipsed and staggered configurations, E/S, to the cooling rate. Theoretical calculations of how the ratio of bandwidth to on-site Coulomb repulsion, $W/U$, are related to E/S through $q(T_g)$ are therefore highly desirable and might render a scaling analysis possible.

\subsection{MIT induced by voltage pulses with increasing amplitude} 
Another way to reach {\it any} possible resistivity state between the metallic (yellow) and the insulating state (green curve), also allowing for an accurate control, is to vary the amplitude of the voltage pulse resulting in a $\rho$ {\it vs.}\ $V$ (resistivity-voltage) measurement, shown in the inset of Fig.\,\ref{fig:01}(b). Here, similar to a 'pump-probe experiment', dc voltage pulses with step-wise increasing amplitudes $V_n$ have been applied for $t = 2$\,s, and a small probing current measures the sample resistivity $\rho_n$ after switching off the voltage and a waiting time of $t_{\rm w} = 20$\,s in each case [as discussed below, see also Fig.\,\ref{fig:02} (inset), the sample is then fully relaxed]. The sketch in the inset of Fig.\,\ref{fig:01}(b) illustrates the procedure, where also the result for $T = 20$\,K with the yellow curve as the initial state (compare purple arrow) is plotted. The resistivity starts to increase at $V_n \gtrsim 3$\,V, shows a hump, which can be associated with the resistivity anomaly at $T_g=75$\,K, and further strongly increases at $V_n \gtrsim 4$\,V with saturation being reached for 5\,V, which means that the sample -- under the actual experimental conditions -- is in the highest possible insulating state. The obvious interpretation is that for  $V_n \gtrsim 5$\,V the increase of sample temperature above $T_g$ leads to the maximum possible cooling rate $q(T = T_g)$ -- given by the coupling to the low-temperature heat bath -- after switching off the voltage pulse.\\  

\subsection{Voltage-controlled cooldown and heat-conduction model} 
\begin{figure}[h]
\includegraphics[width=\columnwidth]{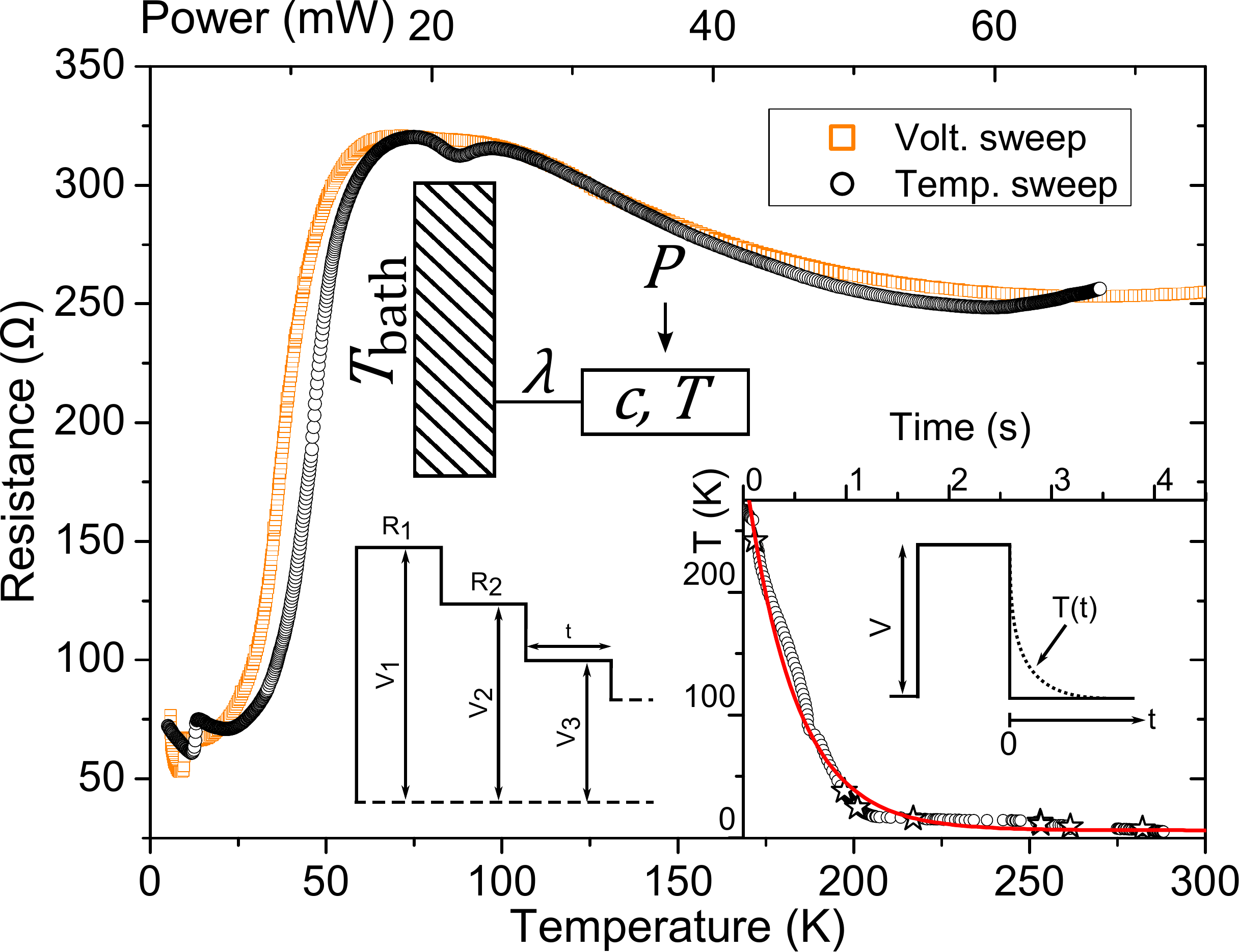} 
\caption{(Color online.) Resistance (2-wire) {\it vs.}\ temperature (black circles) and resistance {\it vs.}\ heating power (orange squares) measurements (see sketch and text) of $\kappa$-(ET)$_2$Cu[N(CN)$_2$]Br. The power can be linked to the temperature axis by the sketched simple model of heat conduction with $T=P/\lambda+T_{\rm bath}$. 
The inset shows the sample temperature {\it vs.} time after switching of a voltage pulse (5\,V applied for $10$\,s), which was extracted using the sample resistance as a reference (see text). Red curve is a fit to the sketched model with $T = T_0 \exp(-\lambda/C \cdot t)+T_{\rm bath}$.}
\label{fig:02}
\end{figure}

In order to corroborate this hypothesis, the following experiment has been performed, without loss of generality, on fully hydrogenated $\kappa$-(ET)$_2$Cu[N(CN)$_2$]Br.
Slow cooling rates controlled by a heating voltage have been realized by carefully ramping down the voltage amplitude in small steps from the saturation heating voltage of $V_n = 5$\,V. As sketched in Fig.\,\ref{fig:02}, after a duration of $t = 3$\,s the amplitude is lowered by $\Delta V=V_n-V_{n+1} = 10$\,mV. At the same time the voltage source-meter allows for measuring the current $I$ through the sample yielding the two-terminal resistance $R$ and the power $P = VI$. In this $R$ {\it vs.}\ $P$ measurement, the temperature-dependent sample resistance has not changed within the 3\,s of the voltage-pulse duration, which confirms the stability of the state that is
determined by the balance of heating (through $P$) and cooling power (through $\lambda$ and $T_{\rm bath}$). A small drift, {\it i.e.} resistance relaxation, however, has been observed (as expected) 
in the temperature range around $T_g$, {\it i.e.} $70 - 80$\,K, where the typical time scale of the structural EEG relaxation is of the same order as the waiting time of a few seconds \cite{Tanatar1999}.

The simple heat-conduction model sketched in Fig.\,\ref{fig:02} implies a power balance of
\begin{equation}
P(t)=C \cdot \dot{T}(t) +\lambda \left[ T(t) - T_{\rm bath} \right],
\label{eq:01}
\end{equation}
where $C$ denotes the sample's specific heat. During the pulse, $\dot{T}(t, P= {\rm const.}) = 0$ and the sample temperature can be extracted from $T= P/\lambda + T_{\rm bath}$. The coupling parameter $\lambda$ follows from the comparison of $R$ {\it vs.}\ $P$ with the temperature-dependent sample resistance $R$ {\it vs.}\ $T$ in a regular resistance measurement in the same configuration, shown in the main panel of Fig.\,\ref{fig:02}. 
Striking features of the sample resistance can be recognized in both experiments, like the minimum at $T \sim 250$\,K ($P \sim 60$\,mW), the maximum at about 75\,K (20\,mW) and the superconducting transition at $T_{c} \sim 12$\,K, ($\sim 5$\,mW). 
Since during the voltage sweep $T_{\rm bath}$ slightly varies and decreases from 8\,K to 5\,K as a function of decreasing $P$, the relation between the sample resistance and the Joule heating power is slightly nonlinear. This correction, however, is neglected in assigning the top axis in Fig.\,\ref{fig:02} to $P$ resulting in a small discrepancy of the curves at low temperatures/heating powers. 
Nevertheless, a striking agreement of the $R(T)$ and $R(P)$ curves within this certainly oversimplified model allows for a crude estimate of the thermal coupling parameter yielding $\lambda \simeq 0.25$\,mW/K. 
%

\subsubsection{Experimental determination of the cooling rate} 
The crucial parameter that determines the physical properties at low temperatures via the strength of electronic correlations $W/U$ is the actual cooling rate at the glass-transition temperature, $q(T_g)$, which is realized after a voltage pulse is switched off.  In order to measure the relaxation of the sample temperature towards $T_{\rm bath}$, we record its resistance as a function of time after the voltage pulse was switched off, see inset of Fig.\,\ref{fig:02}. To that end, $R(t)$ at $T_{\rm bath} = 5$\,K has been measured by initially superimposing the dc heating voltage pulse of $V = 5$\,V with an ac current and detecting the phase-locked ac voltage with a lock-in amplifier, which was read out after the pulse by a fast data acquisition card (National Instruments PCI-6281). A comparison with a regular four-terminal $R(T)$ reference curve (not shown), {\it i.e.}\ essentially using the sample resistance as a thermometer, allows to link the above-mentioned distinct features in both curves, see the star symbols in Fig.\,\ref{fig:02} (inset). 
Solving the simple model of Eq.\,(\ref{eq:01}) yields $T(t) = T_0 \exp(-\lambda/C\cdot t)+T_{\rm bath}$. Crudely neglecting the temperature dependence of the specific heat yields a rough estimate of the cooling rate. The result is shown as a fit to the data in the inset of Fig.\,\ref{fig:02} (red line) yielding a starting temperature $T_0 = 307$\,K, a relaxation time $\tau=C/\lambda=0.5$\,s, a bath temperature $T_{\rm bath}(t)\simeq6$\,K and a cooling rate $\dot{T} \equiv q(T_g) = {\rm d}T/{\rm d}t(75\,{\rm K}) \simeq 138 \pm 30$\,K/s. Despite the large error bar resulting from our oversimplified model, a very fast cooling rate of the order of several 1000\,K/min reproduces our data reasonably well (red line in inset of Fig.\,\ref{fig:02}).\\ 

\section{Relaxation model of glassy dynamics} 
\begin{figure}[ht]
\includegraphics[width=\columnwidth]{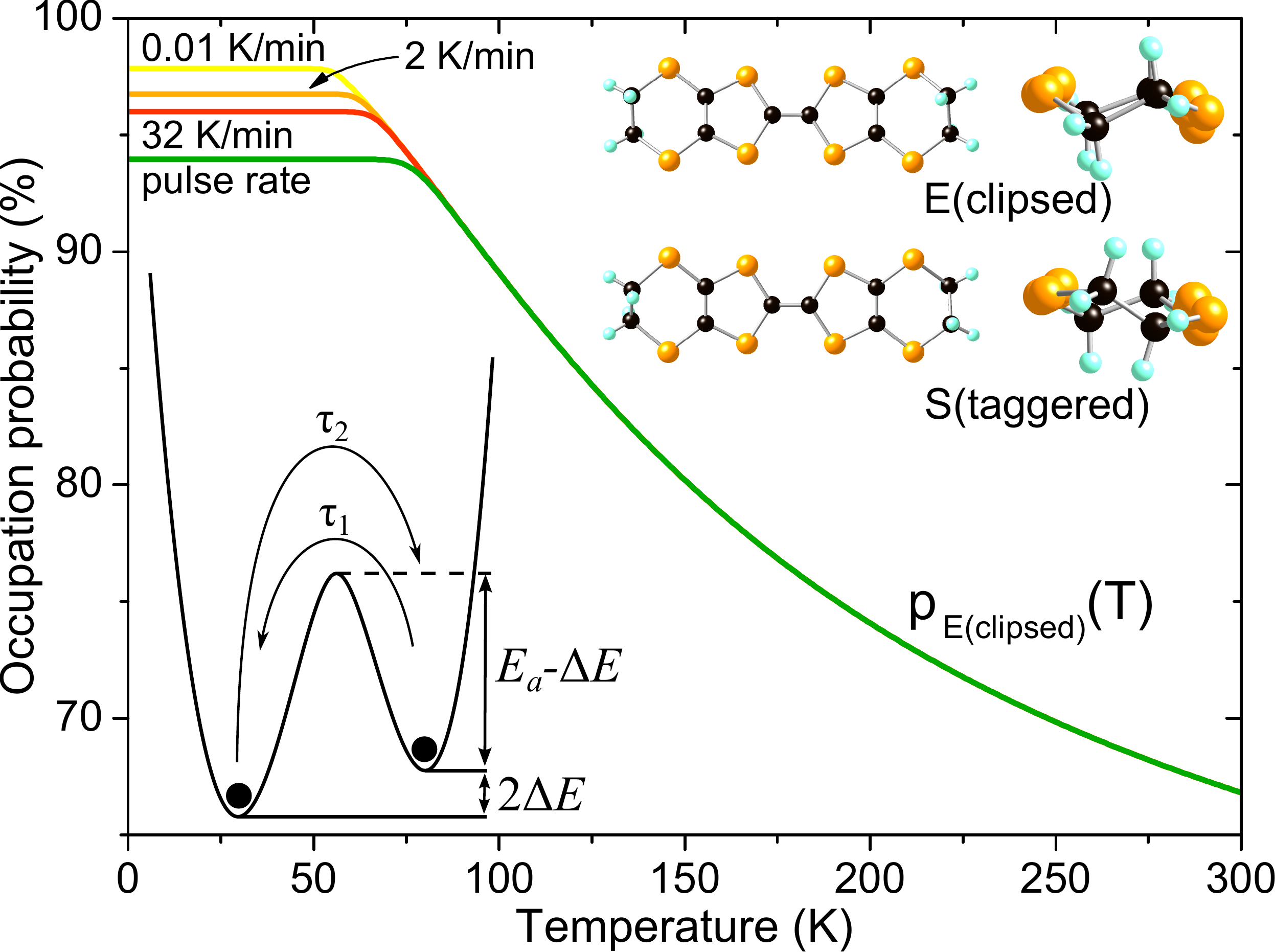} 
\caption{(Color online) Temperature-dependent occupation probability, $p_{\rm E}(T)$, resulting from the general model of Eq.\,(\ref{eq:05}), for the favored eclipsed (E) conformation of the ET molecules' EEG for different cooling rates. "Pulse rate" refers to cooling after a heat pulse is applied as shown in Fig.\,\ref{fig:02} (inset). Faster cooling rates cause the system to fall out of thermal equilibrium at higher temperatures and the relaxation times ($\tau_1$ and $\tau_2$ in the double-well potential) become too large to be observed for temperatures $T < T_g$ resulting in a frozen-in distribution of EEG conformations.}
\label{fig:03}
\end{figure}

After having established the experimental conditions, we discuss the impact of such-obtained 
very fast cooling rates on the relative occupation of the EEG degrees of freedom (E and S, see inset of Fig.\,\ref{fig:03}) affecting the electron-electron correlations in the Mott systems $\kappa$-(BEDT-TTF)$_2$Cu[N(CN)$_2$]Br via the bandwidth, {\it i.e.}\ $W/U$. For this material, the eclipsed conformation E is the lower-energy configuration \cite{Ishiguro1998} and $\tau = \nu_0^{-1} \exp{E_a/k_BT}$ describes the thermally-activated relaxation time with an activation barrier of order $E_a \sim 210 - 260$\,meV as determined from various different experimental techniques for $\kappa$-Br and $\kappa$-Cl, see \cite{Mueller2002} and references therein. At high temperatures, the E and S states are thermally disordered, with preferential population of the E conformation at room temperature, whereas for low temperatures the eclipsed configuration becomes increasingly favorable, until at the glass transition temperature $T_g(q)$ due to kinetic reasons the actual population of E and S becomes frozen. 
\\
%
The occupation probability of the (eclipsed) ground state of the related double-well potential 
in thermal equilibrium is given by
\begin{equation}
p_{\rm E}^\infty(T) = \frac{1}{1+e^{-\frac{2 \Delta E}{k_{\rm B}T}}}. 
\end{equation} 
By quenching the system, a metastable state is realized, which relaxes back to thermal-equilibrium 
with an effective relaxation time
\begin{equation}
\begin{split}
\tau_{\rm eff}&=\left(\frac{1}{\tau_{1}}+\frac{1}{\tau_2}\right)^{-1}\\
&=\nu_0^{-1}{\rm e}^{\frac{E_{a}}{k_{\rm B}T}}\left({\rm e}^{\frac{\Delta E}{k_{\rm B}T}}+{\rm e}^{\frac{-\Delta E}{k_{\rm B}T}}\right)^{-1}
\end{split}
\end{equation}
driven by an attempt frequency $\nu_0$ (usually associated with characteristic phonon frequencies) and determined by the effective energy barrier $E_{a}$  
between the E and S confirmations with an energy difference $2 \Delta E$, see sketch in Fig.\,\ref{fig:03}. 
At $T = T_g \simeq 75$\,K the relaxation time is in the order of typical experimental time scales $\tau_{\rm eff} \sim 100$\,s, which defines the glass transition. 
Thereby, the actual occupation of the EEG orientational degrees of freedom becomes frozen in. In order to quantify this amount of 'disorder' (of the EEG orientations in the ordered crystalline environment) we introduce the time-dependent occupation probability, which describes the relaxation of the system after quench-cooling from $T_0$ to $T_1$:
\begin{equation}
\label{eq:04}
p_{\rm E}(t) = p_{\rm E}^\infty(T_1) - \left[ p_{\rm E}^\infty(T_1) - p_{\rm E}^\infty(T_0) \right] \cdot {\rm e}^{-t/\tau_{\rm eff}}.
\end{equation} 
If instead of an abrupt change from $T_0$ to $T_1$, the temperature changes continuously with cooling rate $q={\rm d}T/{\rm d}t$, the temperature-dependent relaxation is described by expressing Eq.\,(\ref{eq:04}) as a differential equation and a transformation of variables leads to \cite{comment2}
\begin{equation}
\label{eq:05}
\frac{{\rm d}p_{\rm E}(T)}{{\rm d}T} = \frac{p_{\rm E}^\infty(T) - p_E(T)}{q  \tau_{\rm eff}}.
\end{equation}
Figure\,\ref{fig:03} shows the numerical evaluation of this recursive expression for different applied cooling rates $q = 0.01$\,K/min (yellow curve), 2\,K/min, 32\,K/min and for the experimental relaxation curve (shown in the inset of Fig.\,\ref{fig:02}) after switching off a heating pulse (green curve), where $E_{a}/k_B = 2650$\,K has been used for the energy barrier, see \cite{Mueller2002} and references therein. A comparison of our model with the occupation factors for the eclipsed conformation determined from high-resolution synchrotron x-ray diffraction \cite{Wolter2007} yields $\Delta E/k_B = 105 \pm 5$\,K and $\nu_0= 10^{16 (\pm 3)}$\,Hz for the energy difference between the E and S states and the attempt frequency, respectively. Our findings demonstrate that the degree of frozen 'disorder' in terms of the difference in the occupation probabilities of the E and S states remains small for a wide range of cooling rates with $p_{\rm  E} = 97.8$\,\% ($q = 0.01$\,K/min), 96.8,\% (2\,K/min), 96\,\% (32\,K/min) and 94\,\% ("pulse rate", {\it i.e.}\ very fast cooling after voltage heat pulse). Although this amount of randomness in the occupations of the ET molecules' EEG in the eclipsed and staggered conformations may have a considerable impact in the present low-dimensional systems with low carrier concentration, the phenomenology suggests that for moderate cooling rates and in the vicinity of the Mott transition the dominating effect on the electronic properties is a bandwidth-driven correlation effect via the decrease of $W/U$ with increasing $q$.

\section{Summary and Conclusion} 
We have utilized the glass-like structural transition of the ET molecules' EEG ordering in metallic $\kappa$-(ET)$_2$X systems coupled to a low-temperature heat bath in order to induce a Mott metal-insulator transition via a voltage heat pulse. By changing the pulses' amplitude and duration and by exploiting the slow relaxation times around the glass transition temperature $T_g$, essentially any intermediate resistive state between insulating and metallic behavior can be realized if the sample is located close to the Mott transition at a critical value of $W/U$. The application of a simple model, where the eclipsed and staggered conformations of the ET molecules' terminal ethylene groups are described as a double-well potential, to our measurements allows to determine the energy difference $2 \Delta E$ between the E and S state and to quantitatively estimate the amount of EEG units frozen in the non-equilibrium S orientation, as compared to the majority in the E state, for different cooling rates. We find that for moderate and even the rapid cooling rates reported in the literature, the occupation of S conformation is below 5\,\%. This finding agrees very well with theoretical calculations of the conformational disorder of the effective dimer site energy (for holes) affecting the scattering rate of charge carriers in $\kappa$-(ET)$_2$Cu[N(CN)$_2$]Br \cite{Powell2009}. Even for the very large cooling rates of $\sim 1000$\,K/min created in our experiment, the residual S occupation amounts to only 6\,\%, this small value resulting from the relevant energy scales of $E_a \sim 2600$\,K and $\Delta E \sim 100$\,K. The influence of different cooling rates on the electronic ground state properties is caused by two effects: (i) Disorder, {\it i.e.}\ a random lattice potential, the degree of which we have modeled in Fig.\,\ref{fig:03} as a function of cooling rate increases the scattering of charge carriers. (ii) The temperature dependence of the in-plane lattice constants depends on $q$, therefore influencing the bandstructure. From a comparison with the experimental phase diagram of $\kappa$-Cl determined by applying hydrostatic pressure \cite{Lefebvre2000,Kagawa2004}, we roughly estimate that very fast cooling causes a shift in $W/U$ corresponding to about $40$\,bars. If the 'starting position' is close to the critical region $(W/U)_{c}$, a MIT is observed, whereas for the metallic system far away from the Mott transition, the narrowing of the bandwidth doesnÕt affect the ground state properties as extremely as for a system next to the Mott transition, and the more subtle effects of the random lattice potential can explain the experimentally observed suppression in $T_{\rm c}$ of superconducting $\kappa$-Br \cite{Su1998,Powell2009}. For partially deuterated $\kappa$-$d_{0.8}$-Br the strong shift in the phase diagram from the conducting to the insulating side of the Mott MIT across $(W/U)_{c}$, however, very likely is mainly due to the narrowing of the bandwidth (change in $W/U$) caused by the frozen structural configuration which corresponds to a wider lattice as compared to the slowly-cooled situation. Further experimental and theoretical work is highly desirable to discern the role of EEG disorder for random lattice potential and change of $(W/U)$. 
The possibility to quantitatively estimate the degree of EEG disorder outlined here, may be of help for future systematic studies of the electronic properties as a function of cooling rate.\\

We thank the Deutsche Forschungsgemeinschaft (DFG) for financial support within the collaborative research center SFB/TR 49 and acknowledge funding from Grants-in-Aid for Scientific Research from JSPS (25287080). J.M.\ is grateful for fruitful discussions with Peter Lunkenheimer. We acknowledge experimental support from Naoki Yoneyama (Sendai) and Harald Schubert (Frankfurt). 


%

\begin{thebibliography}{}

\bibitem{Kanoda1997} K. Kanoda, Hyperfine Interactions {\bf 104}, 235 (1997).

\bibitem{Imada1998} M. Imada, A. Fujimori, and Y. Tokura, Rev. Mod. Phys. {\bf 70}, 1039 (1998).

\bibitem{Toyota2007} N. Toyota, M. Lang, J. M\"uller, {\it Low-Dimensional Molecular Metals} (Springer, Berlin Heidelberg, 2007).

\bibitem{Lebed2008} A. Lebed (Editor),  {\it The Physics of Organic Superconductors and Conductors}, (Springer Berlin Heidelberg, 2008).

\bibitem{Powell2011} B.J. Powell and R.H. McKenzie, Rep. Prog. Phys. {\bf 74}, 056501 (2011).

\bibitem{Yoneyama2004} N. Yoneyama, T. Sasaki, T. Nishizaki, and N, Kobayashi, J. Phys. Soc. Jpn. {\bf 73}, 184 (2004).

\bibitem{Sasaki2005} T. Sasaki, N. Yoneyama, A. Suzuki, N. Kobayashi, Y. Ikemoto, and H. Kimura, J. Phys. Soc. Jpn. {\bf 74}, 2351 (2005).

\bibitem{Analytis2006} J. G. Analytis, A. Ardavan, S. J. Blundell, R. L. Owen, E. F. Garman, C. Jeynes, and B. J. Powell, Phys. Rev. Lett. {\bf 96}, 177002 (2006).

\bibitem{Powell2009} E. Scriven and B.J. Powell, Phys. Rev. B {\bf 80}, 205107 (2009).

\bibitem{Shinaoka2009a} H. Shinaoka and M. Imada, Phys. Rev. Lett. {\bf 102}, 016404 (2009).

\bibitem{Sano2010} K. Sano, T. Sasaki, N. Yoneyama, and N. Kobayashi, Phys. Rev. Lett. {\bf 104}, 217003 (2010).

\bibitem{Sasaki2012} T. Sasaki, Crystals {\bf 2012}, 2, 274 (2012).


%












\bibitem{Su1998} X. Su, F. Zuo, J. A. Schlueter, M. E. Kelly, and J. M. Williams, Phys. Rev. B  \textbf{57}, R14056 (1998).

\bibitem{Su1998a} X. Su, F. Zuo, J. A. Schlueter, A. M. Kini, and J. M. Williams, Phys. Rev. B  \textbf{58}, R2944 (1998).

\bibitem{Tanatar1999} M.A. Tanatar, T. Ishiguro, T. Kondo, and G. Saito, Phys. Rev. B {\bf 59}, 3841 (1999).

\bibitem{Stalcup1999} T. F. Stalcup, J. S. Brooks, and R. C. Haddon, Phys. Rev. B  \textbf{60} 9309 (1999).

\bibitem{Kawamoto1997} A. Kawamoto, K. Miyagawa, and K. Kanoda, Phys. Rev. B {\bf 55},  14140 (1997).

\bibitem{Taniguchi2003} H. Taniguchi, K. Kanoda, and A. Kawamoto, Phys. Rev. B {\bf 67}, 014510 (2003).

\bibitem{Taylor2008} O. J. Taylor, A. Carrington, and J. A. Schlueter, Phys. Rev. B {\bf 77}, 060503(R) (2008).

\bibitem{Ishiguro1998} T. Ishiguro, K. Yamaji, G. Saito, {\it Organic Superconductors}, Springer 1998 (Berlin, Heidelberg).

\bibitem{Saito1999} K. Saito, H. Akutsu, and M. Sorai, Solid State Commun. {\bf 111}, 471 (1999).

\bibitem{Akutsu2000} H. Akutsu, K. Saito, and M. Sorai, Phys. Rev. B {\bf 61}, 4346 (2000).

\bibitem{Mueller2002} J. M\"uller, M. Lang, F. Steglich, J. A. Schlueter, A.M. Kini, and T. Sasaki, Phys. Rev. B {\bf 65} 144521 (2002).

\bibitem{Mueller2004} J. M\"uller, M. Lang, F. Steglich, and J. A. Schlueter, Journal de Physique IV (France) {\bf 114}, 341 (2004).

\bibitem{comment} The basic concepts of glasses ('classical' glass-forming systems are undercooled liquids), {\it e.g.}, related to the mechanisms of relaxation processes, are rather general in nature and apply also to spin-glasses and {\it glass-like} systems. In the latter, only certain molecular entities show glassy behavior (such that thermodynamic equilibrium cannot be established below $T_g$) in an otherwise crystalline environment.

\bibitem{Mueller2009} J. M\"uller, J. Brandenburg, and J.A. Schlueter, Phys. Rev. B {\bf 79}, 214521 (2009).

\bibitem{Brandenburg2012} J. Brandenburg, J. M\"uller, and J. A. Schlueter, New J. Phys. {\bf 14}, 023033 (2012). 

\bibitem{Powell2004} B. J. Powell and R. H. McKenzie, Phys. Rev. B {\bf 69}, 024519 (2004).

\bibitem{Wang1990} H. H. Wang, A. M. Kini, L.K. Montgomery, U. Geiser, K. D. Carlson, J. M. Williams, J. E. Thompson, D. M. Watkins, and W. K. Kwok, Chem. Mater. {\bf 2}, 482 (1990).

\bibitem{Sasaki2011} T. Sasaki, H. Oizumi, Y. Honda, N. Yoneyama, and N. Kobayashi, J. Phys. Soc. Japan {\bf 80}, 104703 (2011).

\bibitem{Kagawa2005} F. Kagawa, K. Miyagawa, and K. Kanoda, Nature {\bf 436}, 534 (2005).










\bibitem{comment2} Equation (\ref{eq:05}) describes the phenomenon characteristic for glasses that a physical quantity becomes only weakly temperature dependent below $T_g$. It is related to the formally similar equation for the fictive temperature within the Tool-Narayanaswamy theory, see G.W. Scherer, {\it Relaxation in Glass and Composites}, John Wiley \& Sons (1986).

\bibitem{Wolter2007} A. U. B. Wolter, R. Feyerherm, E. Dudzik, S. S\"ullow, Ch. Strack, M. Lang, and D. Schweitzer, Phys. Rev. B {\bf 75}, 104512 (2007).

\bibitem{Lefebvre2000} S. Lefebvre, P. Wzietek,  S. Brown, C. Bourbonnais,  D. J$\acute{{\rm e}}$rome,  C. M$\acute{{\rm e}}$zi$\grave{{\rm e}}$re,  M. Fourmigu$\acute{{\rm e}}$, and P. Batail, Phys. Rev. Lett. {\bf 85}, 5420 (2000).

\bibitem{Kagawa2004} F. Kagawa, T. Itou, K. Miyagawa, and K. Kanoda, Phys. Rev. B {\bf 69}, 064511 (2004).

\end{thebibliography}
%

\end{document}